\documentclass[aps,prl,twocolumn,showpacs,amsmath,amssymb]{revtex4}
\usepackage{epsfig}
\usepackage{bm}
\usepackage{color}

\begin{document}

\title{ Role of Inelastic Tunneling through the Barrier in Scanning Tunneling
Microscope Experiments on Cuprates }

\author{S. Pilgram, T.M. Rice, and M. Sigrist}
\affiliation{Theoretische Physik, ETH Zurich, CH-8093 Z\"urich, Switzerland}

\date\today

\begin{abstract}
The tunneling path between the CuO$_2$-layers in cuprate superconductors and a
scanning tunneling microscope tip passes through a barrier made from other
oxide layers. This opens up the possibility that inelastic processes in the
barrier contribute to the tunneling spectra. Such processes cause one or
possibly more peaks in the second derivative current-voltage spectra displaced
by phonon energies from the density of states singularity associated with
superconductivity. Calculations of inelastic processes generated by apical
O-phonons show good qualitative agreement with recent experiments reported by
Lee et al.~\cite{Lee}. Further tests to discriminate between these inelastic
processes and coupling to planar phonons are proposed.
\end{abstract}
%

%Cuprate superconductors (high-Tc and insulating parent compounds)

%Electronic structure 74.25.Jb,

%Phonons

%Tunneling phenomena; point contacts, weak links, Josephson effects 
%  (for SQUIDs, see 85.25.Dq; for Josephson devices, see 85.25.Cp; 
%  for Josephson junction arrays, see 74.81.Fa)

% Scanning tunneling microscopy (including chemistry induced with STM)

\maketitle

%{\it Introduction} -- 
The importance of the electron-phonon interaction in the high temperature
cuprate superconductors has long been a matter of debate. In conventional
superconductors tunneling experiments analyzed by McMillan and
Rowell~\cite{McMillan} not only unequivocally established that the exchange of
phonons between electrons is the underlying mechanism for Cooper pairing but
also allowed its spectroscopic determination. In the case of cuprates there
are many good reasons to doubt the dominance of an attraction due to phonon
exchange. These reasons range from the d-wave rather than s-wave pairing
symmetry, to their electronic structure as lightly doped Mott insulators. The
latter points toward a strong onsite Coulomb repulsion which cannot be treated
perturbatively, violating a key postulate in the standard BCS-Eliashberg
theory. A number of proposals have been made for a subdominant role for the
electron-phonon interactions particularly involving the half-buckling phonon
modes which couple attractively in the d-wave pairing
channel~\cite{Bulut,Dahm,Nunner,Nazarenko,Jepsen,Honerkamp}.

Very recently, a new set of tunneling measurements at low temperatures has
been reported by Lee et al.~\cite{Lee} using a STM (Scanning Tunneling
Microscope). Results with this local tunneling probe show considerable
variations in the form and energy of the density of states (DOS) singularity
associated with the superconductivity and also a peak in the second derivative
current-voltage spectra which is displaced from the DOS peak by a fairly
constant energy. Lee et al.~\cite{Lee} report further that isotope
substitution on the O-ions shifts this energy consistent with a phonon origin
for this peak. This raises the question whether these spectra should be
interpreted analogously to those of conventional superconductors or if these
peaks have another source.

\begin{figure}[b]
\epsfxsize8.5cm \centerline{\epsffile{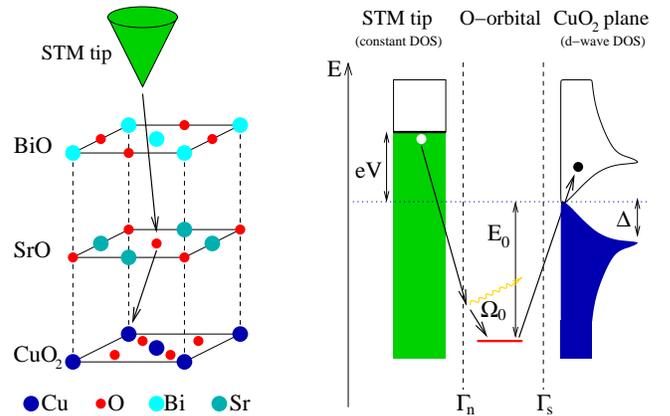}}
\vspace{5mm}
\caption{{\bf Left panel:} Geometry relevant for the scanning tunneling
  microscope experiment, the superconducting CuO$_2$-plane lies below BiO-
  and SrO-layers.  {\bf Right panel:} Suggested inelastic cotunneling process
  via the apical oxygen atom leading to phonon-satellites in the
  current-voltage characteristics in electron-energy scheme. }
\label{Geometry}
\end{figure}

A series of works explain the appearence of side peaks in the DOS of
d-wave superconductors in terms of localized bosonic modes~\cite{Morr},
resonant spin modes~\cite{Balatsky,Zhu1} or the breathing mode
phonons~\cite{Zhu2}. All these proposals consider inelastic processes that
happen directly in the superconducting plane. In this letter we examine an
alternative explanation in terms of inelastic tunneling processes. Such
inelastic processes are well established in electron tunneling
spectroscopy~\cite{Jaklevic,Scalapino,Kirtley1,Kirtley2} of
metal-insulator-metal junctions and in STM spectra taken on metals with
adsorbed molecules~\cite{Molecules,Molecules2}.  In cleaved BSSCO the topmost
layer is the BiO-plane, but the electrons tunnel into and out of the
CuO$_2$-planes. Inelastic tunneling can occur in the intervening barrier,
e.g. when the tunneling hole passes through the apical O-ion. Although the
2p$_z$-orbitals of the apical O-ion do not hybridize with the
3d$_{x^2-y^2}$-orbitals of the Cu-ion directly below there is a weaker
hybridization with orbitals centered on neighboring Cu-ions~\cite{Chen}.
However we will not go into such details here. Rather, we examine a simple
model which has an important tunneling path through an apical O-orbital
removed from the chemical potential. This tunneling path is in turn coupled to
a vibration of this ion. Our aim is to establish the key features of this
inelastic tunneling process and to compare them to the experimental spectra.

%{\it Theory} -- 
Our aim is to calculate the cotunneling current through the
2p$_z$-orbital of the apical O-ion.  For this purpose we describe the setup
depicted in Fig.~\ref{Geometry} by the Hamiltonian,
$$
H = H_{\text{n}} + H_{\text{s}} + H_{\text{t}}
$$
\begin{equation}
+ E_0 \sum_s p_s^\dagger p_s
+ \sum_q \omega_q b_q^\dagger b_q
+ \sum_{qs} M_q p_s^\dagger p_s\left(b_q + b_{-q}^\dagger\right),
\end{equation}
where $H_{\text{n}}$ represents the normal conducting STM tip, $H_{\text{s}}$
the superconducting CuO$_2$-plane and $H_{\text{t}}$ the hopping onto the
apical oxygen orbital. Operator $p^\dagger_s$ creates a hole in the
oxygen p-orbital, operator $b_q^\dagger$ creates a phonon with momentum $q$.
The STM tip is characterized by a constant DOS $N_\text{n}(\varepsilon)
\propto n_\text{n}(\varepsilon)=1$. Partial line widths
$\Gamma_\text{n},\Gamma_\text{s}$ of the 2p$_z$-orbital at energy $E_0$ are
used to describe its hybridization with the STM tip and the superconducting
CuO$_2$-plane. As we shall discuss at the end of this section the above
parameters mainly affect the overall prefactor in the tunneling current and
are not very crucial.  The shape of the current-voltage characteristics is
determined by the DOS of the superconductor and the phonon spectrum.  For the
superconductor we assume for simplicity a d-wave type DOS
$N_\text{s}(\varepsilon) \propto n_\text{s}(\varepsilon) = \langle \text{Re}
\left[ \varepsilon / \sqrt{\varepsilon^2 - \Delta^2\cos^2\theta}\right]
\rangle_\theta$.  However, in the experiment~\cite{Lee} the shape of the
superconducting DOS varies strongly from point to point.  Since the phonon
involves mainly the motion of the apical oxygen atoms we assume an optical
phonon band centered around frequency $\Omega_0$ with a weak dispersion
(i.e. a small band width $W$).

The derivation of the cotunneling current follows closely Ref.~\cite{Sheng}
which deals with a similar non-equilibrium situation in the normal conducting
state. Tunneling is treated to lowest order, while electron-phonon coupling is
treated non-perturbatively allowing for multi-phonon processes. Applying a
Lang-Firsov transformation~\cite{Lang} and standard manipulations of many-body
physics~\cite{Mahan} one arrives at the following expression for the
cotunneling current (we set $\hbar = 1$)
\begin{equation}
I(V) = \frac{e\Gamma_\text{n}\Gamma_\text{s}}{2\pi} \int  \frac{
F_\text{n}(\varepsilon)\bar{F}_\text{s}(\varepsilon)
-
F_\text{s}(\varepsilon)\bar{F}_\text{n}(\varepsilon)}
{\left(\varepsilon-E_0\right)^2 + \left|\Sigma(\varepsilon)\right|^2}
d\varepsilon.
\label{Cotunneling Current}
\end{equation}
The factors $F_x(\varepsilon),\bar{F}_x(\varepsilon)$ with
$x=\text{n},\text{s}$ are obtained from a convolution,
\begin{equation}
\left(
\begin{array}{c}
F_x(\varepsilon)\\
\bar{F}_x(\varepsilon)
\end{array}
\right)
=
\int d\omega n_x(\varepsilon+\omega)
\left(
\begin{array}{c}
B(\omega) f_x(\varepsilon+\omega)\\
B(-\omega) \left[1-f_x(\varepsilon+\omega)\right]
\end{array}
\right),
\label{Modified Occupations}
\end{equation}
of the relevant DOS, $n_x(\varepsilon)$, and the Fermi factors,
$f_x(\varepsilon)$, with the phonon correlation function $B(\omega)$. The
latter will be discussed in detail below. The retarded self-energy
$\Sigma(\varepsilon)$ of the 2p$_z$-orbital reads
\begin{equation}
\Sigma(\varepsilon) = -\frac{i}{2} \sum_{x=\text{n,s}} \Gamma_x 
\left\{ F_x(\varepsilon) + \bar{F}_x(\varepsilon) \right\}.
\end{equation}
If the energy $E_0$ of the virtual oxygen state is the largest energy scale in
the problem, the energy dependence of the denominator of Eq.~(\ref{Cotunneling
Current}) can be neglected~\cite{P Energy}. In this case, we may approximate
the tunneling current by
\begin{equation}
I(V) \sim \int 
\{
F_\text{n}(\varepsilon)\bar{F}_\text{s}(\varepsilon)
-
F_\text{s}(\varepsilon)\bar{F}_\text{n}(\varepsilon)
\} d\varepsilon
\end{equation}
and the parameters $E_0,\Gamma_\text{n},\Gamma_\text{s}$ describing the
virtual state enter only in the (unimportant) prefactor. Even if terms with
$\varepsilon \simeq E_0$ contribute significantly to the
integral~(\ref{Cotunneling Current}), the energy dependence of the denominator
will only lead to an asymmetry of the current-voltage signal on a larger
energy scale.  In the following, we therefore concentrate on the influence of
the phonon spectrum which creates the sharp features visible in the
experiment~\cite{Lee}.

\begin{figure}[t]
\epsfxsize8cm \centerline{\epsffile{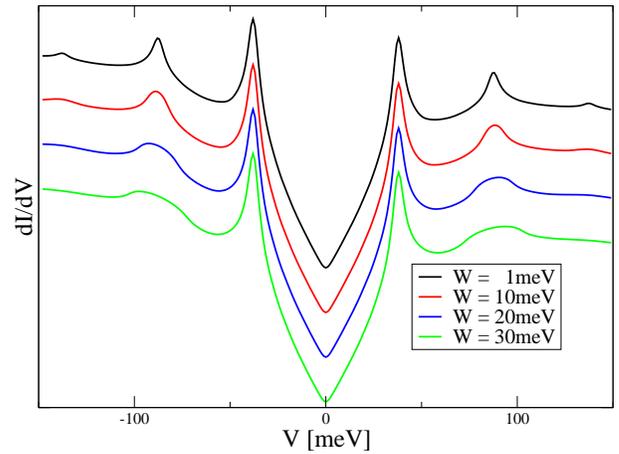}}
\vspace{5mm}
\caption{Shape of $dI/dV$ for different widths $W$ of the phonon
  band. Parameters are $\Delta=38$meV, $\Omega_0=50$meV,$\lambda=0.4$,
  $E_0=2$eV.}
\label{DOS comparison}
\end{figure}

The phonon spectrum enters the current-voltage spectrum (\ref{Cotunneling
Current}) via the modified occupation factors (\ref{Modified Occupations}).
In Eq.\ (\ref{Modified Occupations}), the phonon correlation function
$B(\omega) = (2\pi)^{-1} \int dt e^{i\omega t} 
\langle X(t) X^\dagger(0)\rangle$ describes
possible phonon emission or absorption of the oxygen 2p$_z$-orbital. Those
processes lead to an effective shift of the Fermi occupation factors of both
the superconducting plane and the normal conducting STM tip.  The
``displacement'' operator $X=\exp\left\{ \sum_q M_q (b_q -
b_{-q}^\dagger)/\omega_q\right\}$ results from a Lang-Firsov
transformation~\cite{Lang}.  Replacing the $q$-summation by a frequency
integration over a box density of states and assuming zero temperature for the
phonon bath we obtain the correlation function
\begin{equation}
\langle X(t) X^\dagger(0)\rangle = 
\exp\left\{
\lambda^2 \left(\frac{2}{Wt}\sin\frac{Wt}{2} e^{-i\Omega_0t} -1\right)
\right\}
\label{Phonon Correlator}
\end{equation}
which is used in the following numerical calculations. The parameter $\lambda$
measures the strength of electron-phonon coupling, $\Omega_0$ is the mean
phonon frequency, $W$ defines the band width.  Note that this correlation
function takes into account multiple phonon emissions. A purely perturbative
treatment corresponds to the expansion of the exponential in Eq.\ (\ref{Phonon
Correlator}) to quadratic order in the coupling constant $\lambda$.

%\noindent
%{\it Results} -- 
Fig.\ \ref{DOS comparison} shows the differential tunneling
conductance $dI/dV$ for different widths of the optical phonon band. The
phonon satellites are most clearly visible in the case of completely localized
Einstein phonons ($W=1$meV). Each satellite then reproduces exactly the
singularity of the quasiparticle peak in the superconducting DOS. The case of
Einstein phonons seems to be excluded by the experimental data which shows
broadened phonon satellites. Comparing qualitatively peak heights and peak
widths of theory and experiment, we are led to the conclusion that the phonon
band has a width of roughly $20$meV. This width is probably due to the
vibrational coupling of the apical oxygen atom to its neighbors. Since the
electron-phonon coupling constant is quite large one might also expect
broadening due to particle-hole excitations in the superconducting plane or
due to the anharmonicity of the crystal leading to a finite phonon lifetime.

\begin{figure}[t]
\epsfxsize8cm \centerline{\epsffile{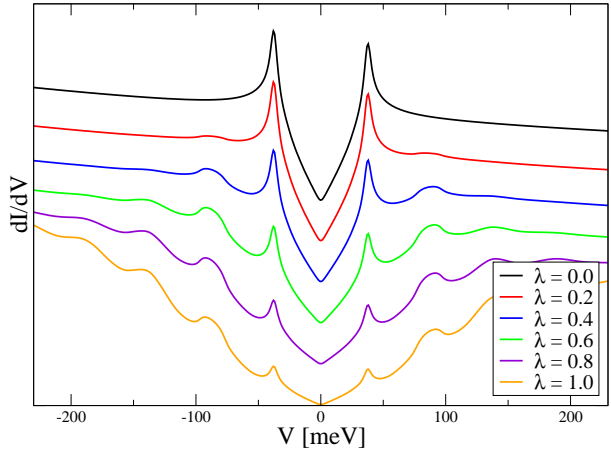}}
\vspace{5mm}
\caption{Shape of $dI/dV$ for different electron-phonon couplings
$\lambda$. Parameters are $\Delta=38$meV, $\Omega_0=50$meV, $W=20$meV,
$E_0=2$eV.}
\label{Coupling comparison}
\end{figure}

In Fig.\ \ref{Coupling comparison} we show the dependence of the differential
conductance on the electron-phonon coupling strength $\lambda$. The stronger
the coupling the more phonon satellites become visible in the spectrum. The
amplitude of the $n$th satellite is smaller than that of the $n-1$th satellite
by roughly a factor of $\lambda^2/n$. A comparison of first and second
satellite therefore allows for an estimate of the coupling strength $\lambda$.
From the experimental data we estimate $\lambda\simeq 0.4$.

\begin{figure}[t]
\epsfxsize8cm \centerline{\epsffile{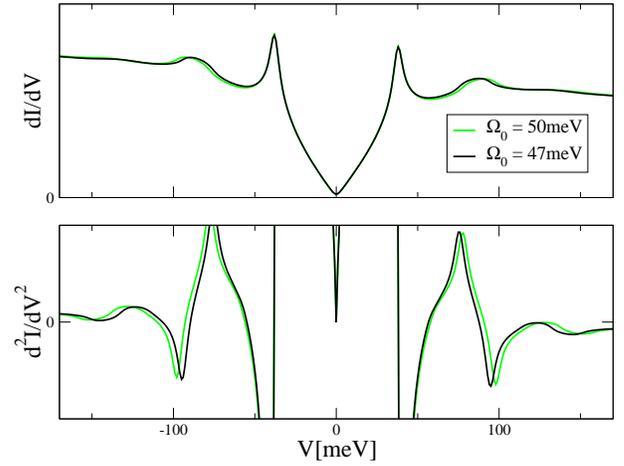}}
\vspace{5mm}
\caption{Isotope shift in the $d^2I/dV^2$.  Parameters are $\Delta=38$meV,
$\Omega_0=50$meV, $W=20$meV, $\lambda=0.4$, $E_0=2$eV. The sign in the
$d^2I/dV^2$ for negative voltages has been flipped.}
\label{Isotope effect}
\end{figure}

%\noindent
Fig.\ \ref{Isotope effect} evaluates the isotope effect in the second
derivative $d^2I/dV^2$ of the current-voltage spectrum as measured in the
experiment~\cite{Lee}. The isotope effect is most clearly visible for the
first satellite which stems from a single convolution of the quasiparticle
peak with the phonon spectrum. The second satellite stems from a twofold
convolution with the phonon spectrum. Its second derivative $d^2I/dV^2$ is
therefore much smoother, and the isotope effect less well seen.

Experimentally~\cite{Lee}, the lowest phonon frequency is obtained as the
energy difference between the quasiparticle peak in the $dI/dV$ (in our case
at 38meV) and the maximum of the second derivative $d^2I/dV^2$ (in our case at
78meV). This procedure is standard in conventional superconductivity where the
strong singularity in the s-wave density of states guarantees a pronounced
feature in the second derivative $d^2I/dV^2$. In the case of d-wave
superconductivity this singularity is much weaker and consequently the maxima
in the second derivative are less well defined. If the quasiparticle peak is
weakened or if the second phonon satellite is investigated another procedure
to determine the phonon frequencies can be preferable: Determining the
positions of both maximum and minimum in the second derivative and taking
their average one can directly estimate the central frequency of the phonon
band.

%{\it Conclusions} -- 
Inelastic tunneling processes show up as a pronounced
peak in the $d^2I/dV^2$ spectrum displaced by a phonon energy from the main
peak in the $dI/dV$ associated with the DOS singularity which accompanies
d-wave superconductivity. The strength and form of the peak reflects those of
the DOS singularity. This feature agrees with experiments of Lee et
al.~\cite{Lee}.  A weaker signal also appears displaced by twice the phonon
energies in agreement with experiment. The phonon energy, 50meV, chosen to
agree with the experimental spectra, is close to the value expected for an
apical O-phonon and rather higher than the energy of the B$_{1g}$
half-buckling mode~\cite{Devereaux}.  One point of divergence between the
theory and the spectra reported by Lee et al. is the presence of a sharp
minimum in the calculated $d^2I/dV^2$ spectra just above the peak.

The STM experiments of Lee et al.~\cite{Lee} show substantial spatial
variations in the spectra with a clear correlation between the frequency shift
of the side peaks and the local value of the superconducting energy gap. Such
a correlation follows if the local bosonic mode couples directly to the planar
holes and if this electron-boson coupling is involved in the
superconductivity. In our model, unlike those of
Refs.~\cite{Morr,Balatsky,Zhu1,Zhu2}, such an explanation does not
apply. However it is known empirically that the positions of the apical O-ions
influences the transition temperature $T_c$ of BSSCO
superconductors~\cite{Eisaki1} and more generally that materials with longer
Cu-apical O bonds have higher values of $T_c$~\cite{Eisaki2}. Therefore the
anticorrelation between the local magnitude of the superconducting gap and the
local size of the frequency shift may reflect this empirical trend and is not
incompatible with our model which ascribes these side peaks to inelastic
tunneling processes in the barrier.

Lee et al.~\cite{Lee} report an isotope shift when all O-sites are occupied by
O$^{16}$ vs. O$^{18}$. A selective isotope replacement only on the apical
O-sites would discriminate between our proposal that the structure is due to
apical O-phonons rather than due to the half-buckling mode involving
displacements of planar O-ions.

In summary we conclude that inelastic tunneling processes involving the apical
O-ions are a viable explanation of the phononic structure in the STM tunneling
spectra recently observed by Lee et al. and we propose selective isotope
substitution as a test of our proposal.

We thank J.~C. Davis, A.~V. Balatsky and the authors of Ref.~\cite{Lee} for
sharing their data with us in advance of publication and F.~C. Zhang,
C. Honerkamp and D.~J. Scalapino for stimulating discussions. We acknowledge
support from the Swiss National Science Foundation and from the MaNEP
(Materials with Novel Electronic Properties) program.


\begin{thebibliography}{99}
\bibitem{P Energy}
The energy $E_0$ needed to create a hole on the 2p$_z$-orbital of the apical
oxygen is typically of the order of $2eV$~\cite{Maekawa}.
\end{thebibliography}
\end{document}